\documentclass[twocolumn]{jsiamletters}
\usepackage{amsmath,amssymb}
\group{Applied Integrable Systems}
\affiliation{Faculty of Fundamental Science and Engineering, Waseda University}{3-4-1 Okubo, Shinjuku-ku, Tokyo 169-8555, Japan}
\affiliation{Faculty of Advanced Science and Technology, Ryukoku University}{1-5 Yokotani, Seta Oe-cho, Otsu, Shiga, 520-2194 Japan}
\affiliation{Faculty of Information Design, Heisei International University}{2000 Mizubuka-Odateno, Kazo-shi, Saitama 347-8504, Japan}
\authorinfo{Daisuke Takahashi}{1*}{daisuket@waseda.jp}
\authorinfo{Junta Matsukidaira}{2}{junta@rins.ryukoku.ac.jp}
\authorinfo{Kazuya Okamoto}{3}{okamoto@hiu.ac.jp}
\email{daisuket@waseda.jp}
\title{Analytical derivation of a three-dimensional fundamental diagram for a real-valued max-plus particle system}
\abstract{
We study a real-valued max-plus particle system on a finite periodic lattice. Its three-dimensional fundamental diagram relates the mean flow to two densities: the conserved particle density and a second density defined by a local state function. We prove that the deviation from the diagram tends to zero for arbitrary initial data with all state values in $[0,1]$. The proof uses two nonnegative local quantities and their evolution relations. For rational initial data with all state values in $[0,1]$, the orbit is eventually periodic, and the fundamental-diagram relation holds exactly on the periodic orbit.}
\makeatletter
\def\ps@arxiv{%
  \let\@oddhead\@empty
  \let\@evenhead\@empty
  \def\@oddfoot{\hfill\footnotesize{\sffamily--\hskip0.5em\thepage\hskip0.5em--}\hfill}%
  \let\@evenfoot\@oddfoot
}
\makeatother

\usepackage{etoolbox}
\makeatletter
\patchcmd{\@maketitle}
  {\small\hskip6.5em\small Received\par}
  {}
  {}
  {\PackageError{arxiv-patch}
    {Failed to remove the Received line}
    {Check the definition of \string\@maketitle\space
     in jsiamletters.cls.}}
\makeatother
\begin{document}
\maketitle
\pagestyle{arxiv}
\thispagestyle{arxiv}
\section{Introduction}

Particle cellular automata (PCA) are conservative cellular automata
 describing interacting particles on a discrete lattice.
Related cellular automaton models reproduce traffic jams
\cite{nagel1992}, and their density--flow relations have been
studied numerically and analytically \cite{schreckenberg1995}.
The evolution of PCA can be written in conservation form with
a local flux function \cite{ref6}. The relation between particle
density and mean flow, called the fundamental diagram,
characterizes their macroscopic behavior.
A typical example is the elementary cellular automaton ECA184,
which is closely related to the ultradiscrete Burgers equation
\cite{ref5}.  For this system, the asymptotic mean flow is uniquely
determined by the particle density.

For particle cellular automata with larger neighborhoods, however,
the situation becomes more complicated.  Several four-neighbor
particle cellular automata have single-valued fundamental diagrams
and their asymptotic behavior can be analyzed by max-plus methods
\cite{ref9}, whereas five-neighbor particle cellular automata
(PCA5) exhibit a much richer variety of asymptotic behavior.
Boccara and Fuk\'s enumerated the binary number-conserving rules
with five neighbors \cite{boccara1998}. Max--min-plus expressions
derived from single-valued fundamental diagrams were subsequently
studied for several classes of PCA5 rules \cite{ref10}.
Other PCA5 rules, however, have multivalued conventional
two-dimensional fundamental diagrams in the density--flow plane:
the same particle density may give different asymptotic mean
flows depending on the initial configuration.
\par
For the rule called PCA5-17R in \cite{okamoto2026},
$\rho_1$ denotes the density of sites in state $1$, and
$\rho_2$ denotes the number of occurrences of the local pattern
$110$ divided by the number of lattice sites.
Both densities are conserved, and the asymptotic mean flow is
uniquely determined by them \cite{ref11}.
Its three-dimensional fundamental diagram consists
of the two planar pieces
\begin{equation}  \label{fund}
             Q=\max(2\rho_2,\,2\rho_1-1).
\end{equation}
Thus, the apparent multivaluedness of the conventional fundamental
diagram can be resolved by introducing an appropriate second
macroscopic quantity.
\par
Recently, this framework has been extended from binary PCA to
real-valued max-plus systems.  In particular, two different
real-valued extensions of PCA5-17R, denoted by mPCA5-17R-1 and
mPCA5-17R-2, were introduced \cite{okamoto2026}.  Numerical experiments showed that,
although their local flux functions and their definitions of the
second density are different, both systems exhibit the same
single-valued three-dimensional fundamental diagram of the same form as (\ref{fund}).
In contrast to the binary PCA5-17R, the second density is
not conserved in general in these real-valued systems.
However, an analytical derivation of this numerically observed
relation remained an open problem.

In this paper, we give an analytical derivation of the
three-dimensional fundamental diagram for mPCA5-17R-2.
We introduce
two local quantities defined as the excess of the actual flux over
the two local branches associated with the planar pieces of the fundamental diagram.  Their evolution
satisfies a simple one-sided relation along a characteristic
direction.  This structure enables us to prove, for arbitrary
initial data with all state values in $[0,1]$, that
\[
 Q^n-\max(2\rho_2^n,\,2\rho_1-1)\longrightarrow0
 \qquad (n\to\infty).
\]
Furthermore, for rational initial data with all state values in $[0,1]$, the orbit eventually becomes
periodic and the above fundamental-diagram relation holds exactly
on the periodic orbit.  Thus, we provide an analytical justification for the numerically observed three-dimensional fundamental diagram of mPCA5-17R-2.
\section{Model and fundamental diagram}

We consider mPCA5-17R-2 on a one-dimensional periodic lattice of $K$
sites.  Let $u_j^n$ denote the state at site $j$ and time $n$, where
the index $j$ is taken modulo $K$.  We take initial data $u^0\in[0,1]^K$.
The local quantity $r_j^n$ and the flux $q_j^n$ are defined by
\begin{align}
 r_j^n&=\min(\max(0,u_j^n+u_{j+1}^n-1),\,1-u_{j+2}^n),
 \label{eq:r}\\
 q_j^n&=\max(u_j^n+u_{j+1}^n-1,\,r_j^n+r_{j+1}^n).
 \label{eq:q}
\end{align}
The time evolution is written in the conservation form as
\begin{equation}
 u_j^{n+1}=u_j^n+q_{j-2}^n-q_{j-1}^n.
 \label{eq:evolution}
\end{equation}
Here $q_j^n$ denotes the flux across the bond from site $j+1$ to site $j+2$.

The interval $[0,1]$ is preserved by the evolution.
To see this, suppose that $0\leq u_j^n\leq1$ for all $j$.
Equation \eqref{eq:r} gives
\[
0\leq r_j^n\leq\min(u_j^n,1-u_{j+2}^n),
\]
and hence \eqref{eq:q} implies
$q_j^n\geq r_j^n+r_{j+1}^n\geq0$.
We first prove the lower bound $u_j^{n+1}\geq0$.
The two arguments of the maximum defining $q_{j-1}^n$
satisfy
\[
\begin{aligned}
& u_{j-1}^n+u_j^n-1
\leq u_j^n\leq q_{j-2}^n+u_j^n,\\
& r_{j-1}^n+r_j^n
\leq r_{j-1}^n+u_j^n
\leq q_{j-2}^n+u_j^n.
\end{aligned}
\]
Thus $q_{j-1}^n\leq q_{j-2}^n+u_j^n$, and
\eqref{eq:evolution} gives $u_j^{n+1}\geq0$.
For the upper bound, the two arguments of the maximum
defining $q_{j-2}^n$ satisfy
\[
\begin{aligned}
&u_{j-2}^n+u_{j-1}^n-1
\leq u_{j-1}^n
\leq q_{j-1}^n+1-u_j^n,\\
&r_{j-2}^n+r_{j-1}^n
\leq 1-u_j^n+r_{j-1}^n
\leq q_{j-1}^n+1-u_j^n.
\end{aligned}
\]
Here we used $q_{j-1}^n\geq u_{j-1}^n+u_j^n-1$
and $q_{j-1}^n\geq r_{j-1}^n$.
Therefore $q_{j-2}^n\leq q_{j-1}^n+1-u_j^n$, and
\eqref{eq:evolution} gives $u_j^{n+1}\leq1$.
By induction, $u^0\in[0,1]^K$ implies
$u^n\in[0,1]^K$ for every $n\geq0$.

\begin{figure*}[t]
 \centering
 \begin{minipage}[b]{0.42\textwidth}
  \centering
  \includegraphics[width=\linewidth,height=5cm,keepaspectratio]{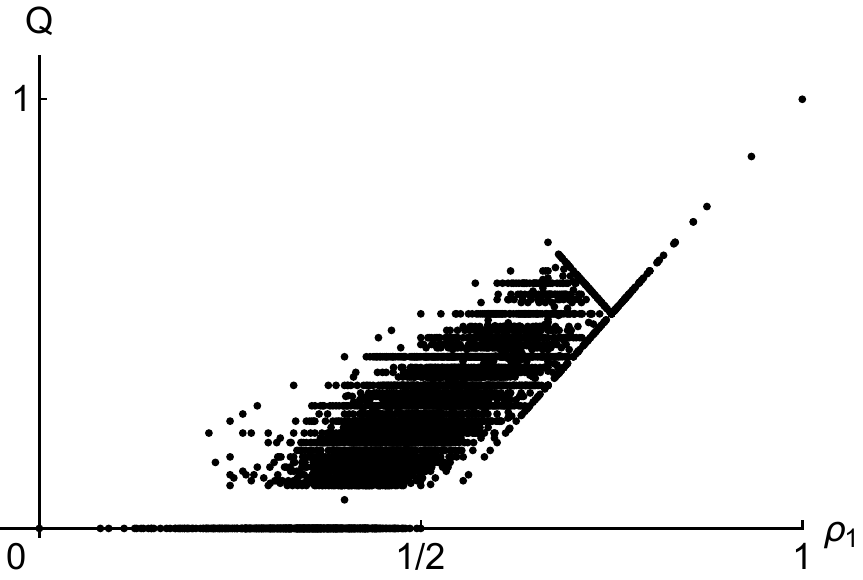}
  \par (a)
 \end{minipage}\hfill
 \begin{minipage}[b]{0.56\textwidth}
  \centering
  \includegraphics[width=\linewidth,height=7cm,keepaspectratio]{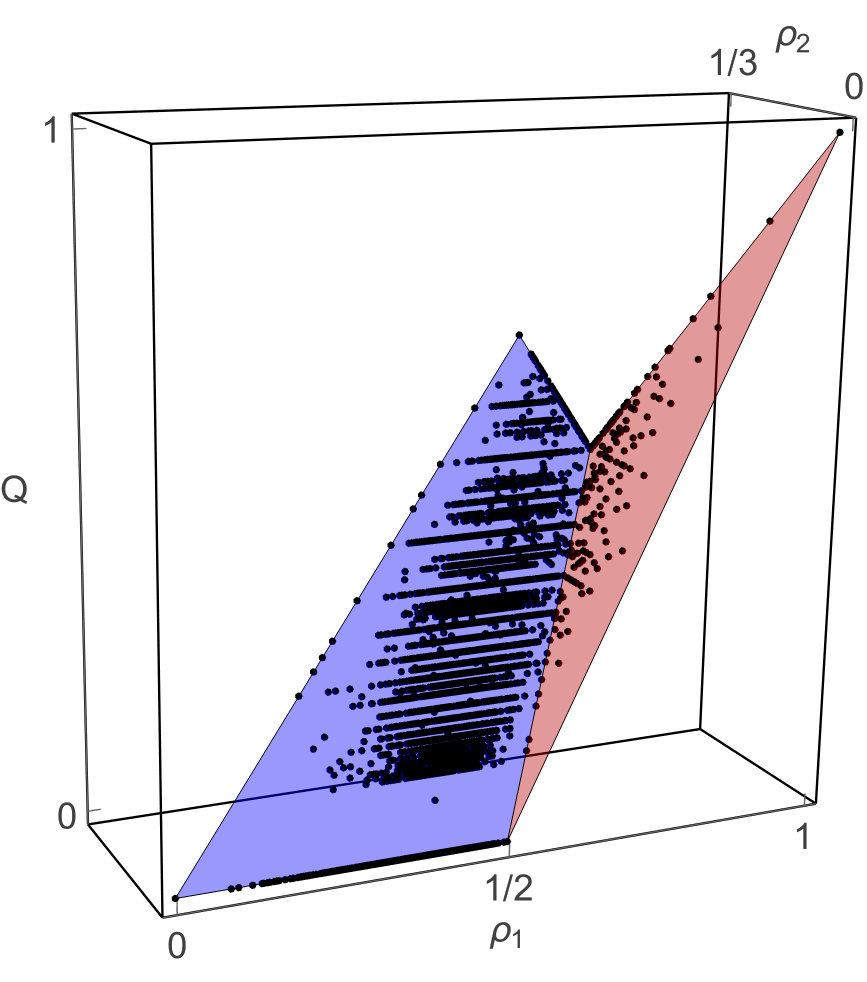}
  \par (b)
 \end{minipage}
\caption{Numerically obtained fundamental diagrams of mPCA5-17R-2.
(a) Conventional two-dimensional diagram using the particle
density $\rho_1$ and mean flow $Q^n$.
(b) Three-dimensional diagram using $\rho_1$, $\rho_2^n$, and $Q^n$.
In each of $10\,000$ trials, the system size $K$ and an integer $M$
were randomly chosen from $\{5,\ldots,20\}$ and $\{1,\ldots,20\}$,
respectively. Each initial state value was randomly chosen from
$\{0,1/M,\ldots,1\}$. Black points show the values at the final
time $n=4K$ of each trial. The colored planar pieces in (b)
represent the surface $Q^n=\max(2\rho_2^n,2\rho_1-1)$
in \eqref{eq:FD}. The superscript $n$ is omitted from the axis labels.}
\label{fig:fundamental-diagrams}
\end{figure*}

We introduce the spatial averages
\begin{equation}
 \rho_1=\frac1K\sum_j u_j^n,\ 
 \rho_2^n=\frac1K\sum_j r_j^n,\ 
 Q^n=\frac1K\sum_j q_j^n.
 \label{eq:rhoQ}
\end{equation}
For binary states, \eqref{eq:r} reduces to
$r_j^n=u_j^n u_{j+1}^n(1-u_{j+2}^n)$, the indicator of the
pattern $110$ at sites $j,j+1,j+2$. Thus $\rho_2^n$ coincides
with the conserved second density $\rho_2$ of the binary system.
Equation \eqref{eq:evolution} immediately shows that $\rho_1$ is
conserved.  Our purpose is to prove that the mean flow approaches
the numerically observed surface,
\begin{equation}
                 Q^n=\max(2\rho_2^n,\,2\rho_1-1).
 \label{eq:FD}
\end{equation}
Figure~\ref{fig:fundamental-diagrams} compares the numerically
obtained two- and three-dimensional fundamental diagrams.
In Fig.~\ref{fig:fundamental-diagrams}(a), the mean flow is not
uniquely determined by the particle density $\rho_1$ alone.
Introducing the second density $\rho_2^n$ resolves this
multivaluedness: the numerical points in
Fig.~\ref{fig:fundamental-diagrams}(b) lie on the surface
\eqref{eq:FD}, shown by the two colored planar pieces.
This observation motivates the analytical derivation below.

For this purpose, we rewrite
\eqref{eq:r}--\eqref{eq:evolution} in a form suitable for the
asymptotic analysis.

\section{Basic relations for the time evolution}

We introduce two local quantities by
\begin{equation}  \label{delta and eps}
 \begin{aligned}
 \delta_j^n&:=q_j^n-(r_j^n+r_{j+1}^n),\\
 \varepsilon_j^n&:=q_j^n-(u_j^n+u_{j+1}^n-1).
 \end{aligned}
\end{equation}
These definitions immediately give
\[
q_j^n=r_j^n+r_{j+1}^n+\delta_j^n,\qquad
q_j^n=u_j^n+u_{j+1}^n-1+\varepsilon_j^n.
\]
Both $\delta_j^n$ and $\varepsilon_j^n$ are nonnegative
by \eqref{eq:q} and \eqref{delta and eps}. They measure the excess of the actual flux
over the two local branches $r_j^n+r_{j+1}^n$ and
$u_j^n+u_{j+1}^n-1$, respectively. These branches represent
local fluxes whose spatial averages give $2\rho_2^n$ and
$2\rho_1-1$, the two branches of the fundamental diagram.
\par
We next derive a local
expression for $\delta_j^n$.

\noindent
{\bf Lemma 1.}
For $0\leq u_j^n\leq1$,
\begin{equation}
 \delta_j^n
 =\max(0,u_j^n+u_{j+1}^n+u_{j+2}^n+u_{j+3}^n-3).
 \label{eq:delta}
\end{equation}

\noindent
{\it Proof.}
Fix $j,n$ and put $S=u_j^n+u_{j+1}^n+u_{j+2}^n+u_{j+3}^n$.
First suppose that $S>3$. Since $u_{j+3}^n\leq1$,
\[
 u_j^n+u_{j+1}^n-1>1-u_{j+2}^n\geq0,
\]
and hence \eqref{eq:r} gives $r_j^n=1-u_{j+2}^n$.
Similarly, we obtain
$r_{j+1}^n=1-u_{j+3}^n$. Consequently,
\[
 r_j^n+r_{j+1}^n=2-u_{j+2}^n-u_{j+3}^n
 <u_j^n+u_{j+1}^n-1.
\]
Equation \eqref{eq:q} gives $q_j^n=u_j^n+u_{j+1}^n-1$,
so the definition of $\delta_j^n$ yields $\delta_j^n=S-3$.

Next suppose that $S\leq3$ and put $A=u_j^n+u_{j+1}^n-1$.
If $A\leq0$, then $r_j^n+r_{j+1}^n\geq0\geq A$.
If $0<A\leq1-u_{j+2}^n$, then $r_j^n=A$ and $r_j^n+r_{j+1}^n\geq A$ holds. It remains to consider $A>1-u_{j+2}^n$.
In this case
\[
 r_j^n=1-u_{j+2}^n,\qquad
 A-r_j^n=u_j^n+u_{j+1}^n+u_{j+2}^n-2>0.
\]
Since $u_j^n\leq1$ and $S\leq3$,
\[
 u_{j+1}^n+u_{j+2}^n-1\geq A-r_j^n,\qquad
 1-u_{j+3}^n\geq A-r_j^n.
\]
Thus \eqref{eq:r} gives $r_{j+1}^n\geq A-r_j^n$.
In all cases $r_j^n+r_{j+1}^n\geq A$, and \eqref{eq:q}
yields $q_j^n=r_j^n+r_{j+1}^n$.
The definition now gives $\delta_j^n=0$.
These two cases prove \eqref{eq:delta}.
\hfill$\Box$

The two local quantities satisfy the following relations, which are
needed to derive the evolution of $\delta_j^n$.

\noindent
{\bf Lemma 2.}
For every $j,n$,
\begin{align}
 \delta_j^n>0
 &\quad\Longrightarrow\quad
 \varepsilon_j^n=0,
 \label{eq:local1}\\
 \delta_j^n=0
 &\quad\Longrightarrow\quad
 u_j^n+u_{j+1}^n+u_{j+2}^n+u_{j+3}^n
 -3+\varepsilon_j^n\leq0.
 \label{eq:local2}
\end{align}

\noindent
{\it Proof.}
Put $A=u_j^n+u_{j+1}^n-1$ and $R=r_j^n+r_{j+1}^n$.
Since $q_j^n=\max(A,R)$, the definitions give
\[
\delta_j^n=\max(0,A-R),\qquad
\varepsilon_j^n=\max(0,R-A).
\]
Hence $\delta_j^n\varepsilon_j^n=0$, proving
\eqref{eq:local1} for arbitrary real state values.

Next suppose that $\delta_j^n=0$. Then $q_j^n=R$, and
\eqref{eq:r} gives
\[
q_j^n=R\leq2-u_{j+2}^n-u_{j+3}^n.
\]
Consequently,
\[
\begin{aligned}
&u_j^n+u_{j+1}^n+u_{j+2}^n+u_{j+3}^n-3
+\varepsilon_j^n\\
&\qquad=q_j^n+u_{j+2}^n+u_{j+3}^n-2\leq0,
\end{aligned}
\]
which proves \eqref{eq:local2}.
\hfill$\Box$

Substituting the identity $q_j^n=u_j^n+u_{j+1}^n-1+\varepsilon_j^n$ into
\eqref{eq:evolution} and shifting the index gives the basic
evolution equation
\begin{equation}
 u_{j+2}^{n+1}
 =
 u_j^n+\varepsilon_j^n-\varepsilon_{j+1}^n.
 \label{eq:uevolution}
\end{equation}

\noindent
{\bf Proposition 3.}
For every $j,n$,
\begin{equation}
 \delta_{j+2}^{n+1}
 =
 \max(0,\delta_j^n-\varepsilon_{j+4}^n),
 \label{eq:deltaevolution}
\end{equation}
and in particular
\begin{equation}
                 \delta_{j+2}^{n+1}\leq\delta_j^n.
 \label{eq:deltamonotone}
\end{equation}

\noindent
{\it Proof.}
Adding \eqref{eq:uevolution} for four consecutive sites yields
\begin{align}
 &u_{j+2}^{n+1}+u_{j+3}^{n+1}
  +u_{j+4}^{n+1}+u_{j+5}^{n+1}-3
 \notag\\
 &\quad =
 u_j^n+u_{j+1}^n+u_{j+2}^n+u_{j+3}^n-3
 +\varepsilon_j^n-\varepsilon_{j+4}^n.
 \label{eq:four}
\end{align}
If $\delta_j^n>0$, Lemma 2 gives
$\varepsilon_j^n=0$; then \eqref{eq:delta} and
\eqref{eq:four} give \eqref{eq:deltaevolution}.

If $\delta_j^n=0$, \eqref{eq:local2} and
$\varepsilon_{j+4}^n\geq0$ show that the right-hand side
of \eqref{eq:four} is nonpositive.  Hence both sides of
\eqref{eq:deltaevolution} are zero.  Finally,
\eqref{eq:deltamonotone} follows from
$\varepsilon_{j+4}^n\geq0$.
\hfill$\Box$

Thus the definitions of $\delta_j^n,\varepsilon_j^n$ and
the evolution relations \eqref{eq:uevolution} and
\eqref{eq:deltaevolution} form the basis of the proof below.

\section{Deviation from the fundamental diagram}

Taking the spatial average of the definition of $\delta_j^n$ gives
\begin{equation}
 Q^n
 =
 2\rho_2^n+\frac1K\sum_j\delta_j^n,
 \label{eq:mean1}
\end{equation}
so that $Q^n\geq2\rho_2^n$.  Similarly, averaging
the identity $q_j^n=u_j^n+u_{j+1}^n-1+\varepsilon_j^n$ gives
\begin{equation}
 Q^n
 =
 2\rho_1-1+\frac1K\sum_j\varepsilon_j^n,
 \label{eq:mean2}
\end{equation}
and hence $Q^n\geq2\rho_1-1$.  Combining the two
identities, we obtain the exact expression
\begin{equation}
 Q^n-\max(2\rho_2^n,2\rho_1-1)
 =
 \frac1K
 \min\left\{
 \sum_j\delta_j^n,\sum_j\varepsilon_j^n
 \right\}.
 \label{eq:deviation}
\end{equation}
Therefore the essential point is to prove that at least one of the
two nonnegative sums on the right-hand side tends to zero.

\section{Initial data in $[0,1]^K$}

\noindent
{\bf Theorem 4.}
For arbitrary real initial data $u^0\in[0,1]^K$,
\begin{equation}
 Q^n-\max(2\rho_2^n,2\rho_1-1)
 \longrightarrow0
 \qquad(n\to\infty).
 \label{eq:main}
\end{equation}

\noindent
{\it Proof.}
Define, only in this proof,
\begin{equation}
                       x_j^n=\delta_{j+2n}^n.
 \label{eq:x}
\end{equation}
By \eqref{eq:deltaevolution},
\begin{equation}
 x_j^{n+1}
 =
 \max(0,x_j^n-\varepsilon_{j+2n+4}^n)
 \leq x_j^n.
 \label{eq:xevolution}
\end{equation}
Thus, for every $j$, $x_j^n$ is nonnegative and nonincreasing,
and therefore $\lim_{n\to\infty}x_j^n$ exists and is nonnegative.

First suppose that
$\lim_{n\to\infty}x_j^n=0$ for every $j$.
For each $n$, the map $j\mapsto j+2n$ is a permutation of the
periodic sites, and hence
$\sum_j\delta_j^n=\sum_jx_j^n.$
Since the sum contains only $K$ terms,
$\sum_j\delta_j^n\to0$.
Equation \eqref{eq:deviation} then proves \eqref{eq:main}.

It remains to consider the case in which
$\lim_{n\to\infty}x_j^n>0$
for some $j$.  Fix such a $j$.  By monotonicity,
$x_j^n\geq\lim_{m\to\infty}x_j^m>0$
for every $n$.  Therefore the positive part in
\eqref{eq:xevolution} never vanishes, and
$\varepsilon_{j+2n+4}^n=x_j^n-x_j^{n+1}.$
The telescoping sum consequently gives
\begin{equation}
 \sum_{n=0}^{\infty}\varepsilon_{j+2n+4}^n
 =
 x_j^0-\lim_{n\to\infty}x_j^n
 <\infty.
 \label{eq:epssum}
\end{equation}

Set $p_n=j+2n+4$, so that $p_{n+1}=p_n+2$.
Equation \eqref{eq:epssum} gives
$\sum_{n=0}^{\infty}\varepsilon_{p_n}^n<\infty.$
We extend this summability to neighboring sites by induction.
Suppose that
\[
 \sum_{n=0}^{\infty}\varepsilon_{p_n+m}^n<\infty
\]
for some $m\geq0$. Applying \eqref{eq:uevolution}
at site $p_n+m$ gives
\begin{equation}
 u_{p_{n+1}+m}^{n+1}
 =
 u_{p_n+m}^n+\varepsilon_{p_n+m}^n
 -\varepsilon_{p_n+m+1}^n.
 \label{eq:propagate}
\end{equation}
Summing over $n=0,\ldots,N-1$, we obtain
\begin{equation}
 \begin{aligned}
 \sum_{n=0}^{N-1}\varepsilon_{p_n+m+1}^n
 &=
 \sum_{n=0}^{N-1}\varepsilon_{p_n+m}^n
 +u_{p_0+m}^0-u_{p_N+m}^N\\
 &\leq
 \sum_{n=0}^{N-1}\varepsilon_{p_n+m}^n+1,
 \end{aligned}
 \label{eq:partial}
\end{equation}
where we used $0\leq u_j^n\leq1$.
By the induction hypothesis, the right-hand side is
bounded above independently of $N$.
Since the left-hand side is a partial sum of nonnegative
terms, it follows that
$\sum_{n=0}^{\infty}\varepsilon_{p_n+m+1}^n<\infty.$
Starting from $m=0$, induction therefore yields
\begin{equation}
 \sum_{n=0}^{\infty}\varepsilon_{p_n+m}^n<\infty,
 \qquad m=0,1,\ldots,K-1.
 \label{eq:allm}
\end{equation}
For each fixed $n$, the $K$ sites
$p_n,p_n+1,\ldots,p_n+K-1$ form the whole periodic lattice.
Hence
\[
 \sum_{n=0}^{\infty}\sum_j\varepsilon_j^n
 =
 \sum_{m=0}^{K-1}
 \sum_{n=0}^{\infty}\varepsilon_{p_n+m}^n
 <\infty.
\]
It follows that
$\sum_j\varepsilon_j^n\longrightarrow0,$
and \eqref{eq:deviation} proves \eqref{eq:main}.
\hfill$\Box$
\section{Rational initial data}

For rational initial data the asymptotic statement can be
strengthened to an exact relation after a finite number of time
steps.

\noindent
{\bf Theorem 5.}
Suppose that $u^0\in[0,1]^K$ and all its components are rational.
Then the orbit enters a periodic orbit after finitely many time
steps, and
\begin{equation}
 Q^n=\max(2\rho_2^n,2\rho_1-1)
 \label{eq:rationalFD}
\end{equation}
holds exactly at every time step on that periodic orbit.

\noindent
{\it Proof.}
Let $M$ be a common denominator of the components of $u^0$.
Equations \eqref{eq:r}--\eqref{eq:evolution} consist only of
maximum, minimum, and additions or subtractions with integer
coefficients.  Therefore every $u_j^n$ remains an integral
multiple of $1/M$.  Since $0\leq u_j^n\leq1$, every component
belongs to the finite set
\[
 \{0,1/M,2/M,\ldots,1\}.
\]
The orbit is confined to a finite set; hence it eventually
becomes periodic.

Let $P$ be its period and put
$D^n=Q^n-\max(2\rho_2^n,2\rho_1-1).$
On the periodic orbit, $D^{n+P}=D^n$, while Theorem 4 gives
$D^n\to0$.  For any time $n$ on the periodic orbit,
$D^n=D^{n+P}=D^{n+2P}=\cdots,$
and the right-hand side tends to zero along this subsequence.
Thus $D^n=0$, proving \eqref{eq:rationalFD}.
\hfill$\Box$

\section{Conclusion}

For initial data in $[0,1]^K$, we have proved convergence of
the deviation from the fundamental diagram \eqref{eq:FD} to zero.
For rational initial data, the orbit is eventually periodic and
the relation holds exactly on the periodic orbit.

Our result does not require convergence of the second density
$\rho_2^n$. Numerical experiments suggest that $\rho_2^n$ converges
to a constant for every initial state in $[0,1]^K$, but a proof
of this conjecture remains an open problem.

Numerical experiments also suggest that \eqref{eq:FD} holds
asymptotically for initial data in $\mathbb{R}^K$.
Establishing this extension beyond $[0,1]^K$, and extending the
present approach to other real-valued particle systems, remain
subjects for future work.

\end{document}